\documentclass[aps,prc,preprint,groupedaddress]{revtex4}

\newcommand{\beq}{\begin{equation}}
\newcommand{\eeq}{\end{equation}}
\newcommand{\beqa}{\begin{eqnarray}}
\newcommand{\eeqa}{\end{eqnarray}}
\usepackage{epsfig}
\usepackage{youngtab}

\begin{document}

 \title{Five-quark components in $\Delta(1232)\rightarrow N\pi$ decay}

\author{Q. B. Li}
\email[]{ligb@pcu.helsinki.fi}
\affiliation{Helsinki Institute of Physics
POB 64, 00014 University of Helsinki, Finland}

\author{D. O. Riska}
\email[]{riska@pcu.helsinki.fi}
\affiliation{Helsinki Institute of Physics
and Department of Physical Sciences, POB 64,
00014 University of Helsinki, Finland}

\thispagestyle{empty}

\date{\today}
\begin{abstract}
Five-quark $qqqq\bar q$ components in the $\Delta(1232)$ are 
shown to contribute significantly to 
$\Delta(1232)\rightarrow N\pi$
decay through quark-antiquark annihilation transitions.
These involve the overlap between the $qqq$ and the
$qqqq\bar q$ components and may be
triggered by the confining
interaction 
between the quarks. With a $\sim$ 10\% admixture
of five-quark components in  
the $\Delta(1232)$
the decay width can be larger by factors $\sim$ 2 -- 3
over that calculated in the
non-relativistic quark model with 3 valence quarks, depending on the
details of the confining interaction.
The effect of transitions between the $qqqq\bar q$ components
themselves on the calculated
decay width is however small.
The large contribution of the quark-antiquark 
annihilation transitions
thus may compensate the underprediction
of the width of the $\Delta(1232)$ by the valence quark
model, once the $\Delta(1232)$ contains $qqqq\bar q$
components with $\sim$ 10\% probability.

\end{abstract}


\maketitle

\section{Introduction}

While the constituent valence quark model provides a simple
and almost quantitative phenomenological description of the
magnetic moments of the octet baryons, it does in its simplest 
versions, where the pions couple 
directly to the quarks,
lead to values for the decay width of the $\Delta(1232)$,
which are only about one half of the empirical value. More
sophisticated covariant versions of the quark model with
realistic wave functions for the 3-quark system fail to
overcome this deficiency \cite{bruno,melde}. Coupled
channel treatments of the hadronic interacting
$\pi N \Delta$ system suggest that
the problem may be cured by the ``pion cloud'' contribution,
which is automatically included in that approach \cite{sato,sato2}.

Here this question is addressed by an
extension of the non-relativistic valence
quark model to include explicitly those 5-quark $qqqq\bar q$ 
configurations in the
proton and the $\Delta(1232)$, which are expected to have
the lowest energy. The presence of such ``sea-quark''
contributions in the proton has been demonstrated
 in several experiments
\cite{towell,NA51,NMC,HERMES}. Given the presence of
explicit $qqqq\bar q$ components in both the nucleon and
(the expected presence in)
the $\Delta(1232)$ resonance, pion decay of the
latter may take place in the form of transitions 
between the respective $qqqq\bar q$ components but
also as annihilation transitions of the form
$qqqq\bar q\rightarrow qqq\pi$ in addition to the
conventional quark model transitions between the
pure $qqq$ states. Here both cases are considered, with
the result that only the annihilation transitions 
contribute significantly.
The amplitude for these
are - at least on the basis of
a qualitative estimate based on simple wave function
models - strong enough to increase the 
calculated width in the quark
model by factors 2 -- 3 if there is a $10$\% probability for
$qqqq\bar q$ components in the 
the $\Delta(1232)$, the magnitude depending on the
model for the confining interaction. The
transitions between the explicit $qqqq\bar q$ components
themselves are however found to be but of minor significance,
mainly because of the small amplitude of the $qqqq\bar q$
component in the proton.   

In section \ref{sec1} the $qqqq\bar q$ configurations in
the nucleon and the $\Delta(1232)$, which are expected
to have the lowest energy, and the pionic
transitions between these are considered. In section
\ref{sec2} the $qqqq\bar q\rightarrow qqq\pi$ transitions
are treated along with a numerical estimate of their
significance. Finally 
section \ref{disc} contains a summarizing
discussion.

\section{Five-quark components in the proton and the $\Delta(1232)$}
\label{sec1}

\subsection{Low lying $qqqq\bar q$ configurations}
\label{sec11}

Positive parity demands that in a $qqqq\bar q$
component in a baryon either one of the 4 quarks or the
antiquark $\bar q$ is orbitally excited to the
$P-$shell. If the 4 quarks are in the ground state, the
corresponding spatial wave function is completely
symmetric, $[4]_X$, and overall antisymmetry demands
that the flavor-spin state have mixed symmetry
$[31]_{FS}$, which can combine with the color state
with the conjugate mixed symmetry $[211]_C$ to total antisymmetry
$[1^4]$. If the antiquark is in its ground state,
the spatial state of the orbitally excited $qqqq$ system
has to have the mixed symmetry
$[31]_X$. In this case overall antisymmetry allows the
flavor-spin state to have the following symmetries:
$[4]_{FS}$, $[31]_{FS}$ or $[211]_{FS}$. The possible
symmetry configurations of the $qqqq$ system that meets
these conditions have been classified in ref. \cite{helminen}. 
The $qqqq\bar q$ configurations that are most likely
to have appreciable probabilities in the proton and the
$\Delta(1232)$ are those, which have the lowest energy
and (or) the strongest coupling to the main $qqq$ 
configuration.

The energy levels of these $qqqq\bar q$ configurations
are split by the hyperfine interaction between the 
quarks. The configuration with the lowest energy depends
on the form of this interaction. 
If the hyperfine interaction is spin-dependent, as usually
assumed, the $qqqq$ configurations
that have the lowest energy are those with the most antisymmetric
spin state, which is the the mixed symmetry state $[22]$.
This is the case if the hyperfine interaction is described
by the colormagnetic interaction
and also if the interaction 
is described by the schematic flavor and spin dependent
interaction $-C_\chi\sum_{i<j} \vec\lambda_F^i\cdot
\vec\lambda_F^j\,\vec\sigma_i\cdot\vec\sigma_j$, which
leads to the empirical ordering of the baryon spectrum 
if $C_\chi\sim 20-30$ MeV \cite{glozman}. In both cases
the flavor-spin symmetry of the $qqqq$ part of the
the lowest energy $qqqq\bar q$
component in the proton is likely to be
$[4]_{FS}[22]_F[22]_S$. 
Because the total isospin of this
$qqqq$ configuration is 0, it cannot be a component in
the $\Delta(1232)$, however. The lowest energy $qqqq$
configuration in the $\Delta$ has the flavor-spin
symmetry $[4]_{FS}[31]_F[31]_S$. In these configurations
the antiquark $\bar q$ is in its ground state. The
other $qqqq\bar q$ configurations in the proton
and the $\Delta(1232)$ are expected to have a much higher
energy \cite{helminen}.

\subsection{Transitions between five-quark configurations}
\label{sec12}

In the ``chiral quark'' model pions couple directly to
constituent quarks. The transition operator for a
transition of the type $\Delta^{++}(s_z = 3/2)
\rightarrow p(s_z =1/2)\pi^+$ is then in the non-relativistic
approximation:
\begin{equation}
T_\pi = -i{g_A^q\over f_\pi} \sum_i 
\tau_- ^i 
\sigma_- ^i\,q_{\pi +}\, .
\label{trans}
\end{equation}
Here the sum runs over the quarks and $g_A^q$ and
$f_\pi$ are the axial vector coupling constant of
the constituent quarks, and $f_\pi$ is the pion decay
constant. The pion momentum component $q_{\pi +}$
is defined as $q_{\pi +}
= -(q_{\pi x}+i q_{\pi y})/\sqrt{2}$.

The matrix element of (\ref{trans}) in the valence
quark model with conventional 3-quark flavor and
spin wave functions \cite{close} for the transition
$\Delta^{++}(s_z=3/2)
\rightarrow p(s_z=1/2)\pi^+$ is:
\begin{equation}
\langle p,1/2\vert T_\pi\vert \Delta^{++},3/2\rangle = 
-i{g_A^q\over f_\pi}\sqrt{2}q_{\pi +}\,(1-{\vec q\,^2\over
6 \omega_3^2})\, .
\label{3qm}
\end{equation}
The last factor accounts for the spatial
extent of the $qqq$ component of the baryon in the
harmonic oscillator model. The parameter $\omega_3$
may be determined from the empirical radius of the
proton as $\omega_3 = 1/r_p \simeq 225$ MeV.  

Consider then the corresponding matrix element for the
$qqqq\bar q$ components in the
proton and the $\Delta(1232)$. If the amplitudes of the
$[4]_{FS}[22]_F[22]_S$ component of the proton is denoted
$A_{p5}$, the corresponding wave function is
\begin{eqnarray}
&&\psi_p(s_z=1/2) ={A_{p5}\over \sqrt{6}}\sum_{a,b}
\sum_{m,s} (1,1/2,m,s\vert\, 1/2,1/2)\,
\nonumber\\
&&[211]_C(a)\,[31]_{X,m}(a)\, [22]_F(b)\,[22]_S(b)\, \bar\chi_s\,
\varphi(\{r_i\})\, .
\label{5qp}
\end{eqnarray}
Here the color, space and flavor-spin wave functions of the
$qqqq$ subsystem have been denoted by their Young patterns
respectively, and the sum over $a$ runs over the 3 
configurations of the $[211]_C$ and $[31]_X$ representations
of $S_4$, and the sum over $b$ runs over the 2 
configurations of the
$[22]$ representation of $S_4$ respectively \cite{chen}.  
Note that as the isospin of the $qqqq$ of the $[22]_F$
configuration is 0, the antiquark can only be a $\bar d$
quark.

The wave function for the $[4]_{FS}[31]_F[31]_S$ configuration
in the $\Delta^{++}$ is

\begin{eqnarray}
&&\psi_\Delta(3/2)^J ={A_{\Delta 5}^{(J)}\over 3}\sum_{a,b}
\sum_{m,s,M,j}
(1,1,m,M\vert\, J,j)(J,1/2,j,s\vert\, 3/2,3/2)\,
\nonumber\\
&&[211]_C(a)\,[31]_{X,m}(a)\, [31]_F(b)\,[31]_{S,M}(b)\, \bar\chi_s\,
\varphi(\{r_i\})\, .
\label{q5d}
\end{eqnarray}
Here $J$ denotes the total angular momentum of the
$qqqq$ system, which takes the values 1 and 2,
and $A_{\Delta 5}^{(J)}$ is the amplitude
of the configuration in the $\Delta(1232)$. 
The sum over $a$ again runs over the 3 
configurations of the $[211]_C$ and $[31]_X$ representations
of $S_4$. Here the sum over $b$ runs over the 3 
configurations of the
$[31]$ representation.  

It is then a straightforward task to calculate the
ratio of the matrix element of the
operator (2) for $\Delta^{++}_{3/2}
\rightarrow p_{1/2}\pi^+$ process in the $qqqq\bar q$
configuration to that in the conventional
$qqq$ configuration. The result is:
\begin{equation}
{_5\langle p,1/2 \vert T_\pi\vert \Delta^{++},3/2\rangle_5
\over _3\langle p,1/2\vert T_\pi\vert \Delta^{++},3/2\rangle_3}
={\sqrt{2}\over 3}{A_{p5}\over A_{p3}A_{\Delta 3}}
(A_{\Delta 5}^{(1)} + \sqrt{5} A_{\Delta 5}^{(2)})\, . 
\label{ratio1}
\end{equation} 
Here $A_{p3}$ and $A_{\Delta 3}$ are the amplitudes for the
$qqq$ component in the proton and the $\Delta(1232)$, 
respectively.
In this expression one should in principle also include
the ratio of the momentum dependent factors that 
account for the spatial extent of the baryon. In the 
harmonic oscillator model these
factors are $1-\vec q\,^2/6\omega_3^2$ for the
$qqq$ configuration and $1-\vec q\,^2/5\omega_5^2$ for
the $qqqq\bar q$ configuration. If the spatial extent
of the 3- and the 5-quark components is the same, so that
\begin{equation}
\omega_5 = \sqrt{{6\over 5}}\omega_3\, ,
\label{omega}
\end{equation}
this ratio is 1. The relative magnitude of the (inverse)
size parameters $\omega_3$ and $\omega_5$ will depend
on the interaction that couples the $qqq$ and the
$qqqq\bar q$ components.

The relative difference from the valence quark model result
that inclusion of the $qqqq\bar q$ configurations brings,
is obtained by addition of the product of the amplitudes
for the $qqq$ components $A_{p5}A_{\Delta 3}$ to the
ratio (\ref{ratio1}):
\begin{equation}
\delta = A_{p3}A_{\Delta 3}(1 +
{\sqrt{2}\over 3}{A_{p5}\over A_{p3}A_{\Delta 3}}
(A_{\Delta 5}^{(1)} + \sqrt{5} A_{\Delta 5}^{(2)}))\, . 
\label{5qratio}
\end{equation}

While the flavor-spin $qqqq\bar q$ component 
$[4]_{FS}[22]_F [22]_S$ 
does not contribute to the magnetic moment of the proton,
it does contribute an amount $A_{p5}^2/3\, \mu_N$ to that of 
the neutron. The ratio of the proton to the 
neutron magnetic moment in the $qqq$ configuration is
-$3/2$ and thus close to the empirical ratio -1.46 
in the static quark model. In covariant versions of the
valence quark model the calculated ratio varies between
-1.46 and -1.66 \cite{brunomag}.  
A large value for  
the amplitude $A_{p5}$ for the $qqqq\bar q$ component
would bring large deviations from the empirical magnetic
moment ratio.

The expression (\ref{5qratio}) reveals that the $qqqq\bar q$
components only in the case $J=2$ can lead to an increase
of the calculated decay rate of $\Delta^{++}\rightarrow p\pi^+$
only when the probabilities of the $qqqq\bar q$ components
are fairly large.
As an example consider the case in which the $qqqq\bar q$
probabilities of the component of the proton and 
the $\Delta^{++}$ (with $J=2$) are 10\% and 
20\%, respectively. In this case $A_{p3}=0.95$ and
$A_{\Delta 5}^{(2)} = 0.45$ 
and $\delta=0.997$ so that there is no net
enhancement. If the $qqqq\bar q$ component of the
proton would be as large as 20\% there would be
a net enhancement of 2 \%.
The conclusion then follows that
the transitions between the
$qqqq\bar q$ components do at most lead to an enhancement
of the calculated decay width by a few percent. 

The proton may also have an admixture with the 
flavor-spin symmetry structure $[4]_{FS}[31]_F[31]_S$, in
which case the antiquark may be either a $\bar u$ or
$\bar d$. The empirical evidence for the flavor
asymmetry of the $q\bar q$ components in the proton
\cite{garvey} 
suggests that this should have a smaller probability
than the component with flavor-spin symmetry
$[4]_{FS}[22]_F[22]_S$. This is also consistent 
with the fact that it is energetically less favorable
\cite{helminen}. The corresponding proton wave function
has the form
\begin{eqnarray}
&&\psi_p(1/2) ={A_{p5}^{(J)}\over 3}\sum_{a,b}
\sum_{m,s,M,j}
(1,1,m,M\vert\, J,j)(J,1/2,j,s\vert\, 1/2,1/2)\,
\nonumber\\
&&(1,1/2,T,t\vert 1/2,1/2)\,
[211]_C(a)\,[31]_{X,m}(a)\, [31]_{F,T}(b)\,[31]_{S,M}(b)
\, \bar\chi_{t,s}\,
\varphi(\{r_i\})\, .
\label{p52}
\end{eqnarray}
Here the isospin-z component of the 4-quark state is
denoted $T$ and that of the antiquark $t$. 
In this configuration the ratio of the amplitudes for
the $\Delta^{++}_{3/2}
\rightarrow p_{1/2}\pi^+$ process in the $qqqq\bar q$
configuration to that in the conventional
$qqq$ configuration is:
\begin{equation}
{_5\langle p,1/2 \vert T_\pi\vert \Delta^{++},3/2\rangle_5
\over _3\langle p,1/2\vert T_\pi\vert \Delta^{++},3/2\rangle_3}
=-{\sqrt{2}\over 6}({A_{p5}^{(1)} A_{\Delta 5}^{(1)}
\over A_{p3}A_{\Delta 3}}+2\sqrt{2}{A_{p5}^{(0)} 
A_{\Delta 5}^{(1)}
\over A_{p3}A_{\Delta 3}}+\sqrt{5}{A_{p5}^{(1)} 
A_{\Delta 5}^{(2)}
\over A_{p3}A_{\Delta 3}})
\label{ratio2}
\end{equation} 
As the magnitude of the numerical coefficients on the 
right hand side of this expression is less than 1 and the
sign of the ratio is negative,
the transitions
between these $qqqq\bar q$ components 
cannot increase the
calculated decay rate for $\Delta^{++}\rightarrow p\pi^+$
over the result obtained in the pure $qqq$ model calculation.

\section{Five-quark to three-quark transitions}
\label{sec2}

\subsection{Direct quark-antiquark annihilation}

\begin{figure}[t]
\vspace{20pt} 
\begin{center}
\mbox{\epsfig{file=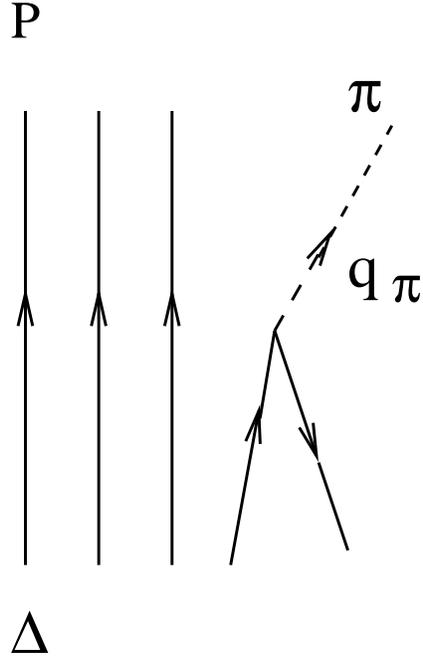, width=55mm}} 
%\mbox{\epsfig{file=f1,width=55mm}} 
\caption{Direct $qqqq\bar q\rightarrow qqq\pi$ annihilation
diagram}
\label{fig1}
\end{center}
\vspace{10pt} 
\end{figure}

The simplest $qqqq\bar q \rightarrow qqq + \pi$
decay mechanism that can contribute to the decay
of the $\Delta(1232)$ is $q\bar q\rightarrow \pi$
pair annihilation process in Fig.\ref{fig1}. The corresponding
amplitude is
\begin{equation}
T_\pi = i\sqrt{2}{m_q g_A^q\over f_\pi}\bar
v(p_{\bar q})\gamma_5 u(p_q).
\label{annix}
\end{equation} 
Calculation of the matrix element of this amplitude
for the decay $\Delta^{++}_{3/2}\rightarrow p_{1/2}\,\pi^+$
requires the calculation of the overlap of the
$qqq$ component of the proton with the residual
$qqq$ component that is left in the $\Delta^{++}$
after the annihilation of a $u\bar d$ pair. It
also requires a specification of the spatial
part of the $\Delta^{++}$ wave function.

The spatial wave function may be expressed with the
help of the following relative coordinates:
 \begin{eqnarray}
&&\vec \xi_1 ={1\over\sqrt{2}}(\vec r_1 -\vec r_2)\, ,\quad
\vec\xi_2 = {1\over\sqrt{6}}(\vec r_1 +\vec r_2 -2 \vec r_3)\, ,
\nonumber\\
&&\vec \xi_3 ={1\over\sqrt{12}}(\vec r_1 +\vec r_2 +\vec r_3
 - 3\vec r_4)\, ,\nonumber\\
&&\vec\xi_4 = {1\over\sqrt{20}}(\vec r_1 +\vec r_2 + \vec r_3
+\vec r_4 - 4 \vec r_5)\, .
\label{relative}
\end{eqnarray}
Here $\vec r_i$ represents the position operator of the
i:th constituent. To form a complete set of basis
vectors, the set (\ref{relative}) may be completed
with the center-of-mass vector $\vec R =
\sum_i \vec r_i/\sqrt{5}$.
The three components of the spatial
state with $[31]_X$ mixed symmetry
may be formed as normalized combinations of a spatially
symmetric function that is multiplied by the 
vectors $\vec \xi_1$, $\vec \xi_2$ and $\vec \xi_3$
respectively, times a completely symmetric function
of the coordinates.

For the present purposes it suffices to describe the
completely symmetric function of the 4 quark coordinates
as a product of harmonic oscillator functions:
\begin{equation}
\varphi(\xi_i) = ({\omega_5^2\over\pi})^{3/4}\,
e^{-\xi_i^2\,\omega_5^2/2}\, ,
\label{harmosc}
\end{equation}
where $\omega_5$ is the constant parameter (6) and $i=1,2,3$.
A similar oscillator wave function is employed for
the antiquark.

The desired annihilation matrix element will take the form
\begin{eqnarray} 
&&\langle T \rangle = 4 A_{p3}A_{\Delta 5}^{(J)}
\,\int \Pi_{i=1}^5 d^3 r_i\,
\psi_{p3}(\vec r_1,\vec r_2,\vec r_3)\,e^{i\vec q_\pi\cdot
(\vec r_4+\vec r_5)/2}\,\nonumber\\
&&\delta(\vec r_4-\vec r_5)\,
\langle T_{45}\rangle \,
\psi_{\Delta 5}(\vec r_1,\vec r_2,
\vec r_3,\vec r_4,\vec r_5) \delta(\vec R)\, .
\label{annihi}
\end{eqnarray}
Here the pure three quark proton wave function is denoted
$\psi_{p3}$ and the coordinates of the annihilated quark and
antiquark are taken to be $\vec r_4$ and $\vec r_5$. 
The matrix element of the annihilation amplitude (\ref{annix})
for annihilation of the 4th quark and the 
antiquark (with coordinate $\vec r_5$) is denoted 
$\langle T_{45}\rangle $.
It is advantageous to express the matrix element in terms
of the relative coordinates:
\begin{eqnarray} 
&&\langle T \rangle = 4
({2\over \sqrt{5}})^3 
A_{p3}A_{\Delta 5}^{(J)}\,\int \Pi_{i=1}^4 d^3 \xi_i\,
\psi_{p3}(\vec \xi_1,\vec \xi_2)\,e^{-i2\sqrt{3}\vec q_\pi\cdot
\vec \xi_3/5}\,\nonumber\\
&&\delta(\vec \xi_4-\sqrt{{3\over 5}}\xi_3)\,
\langle T_{45}\rangle\,
\psi_{\Delta 5}(\vec \xi_1,\vec \xi_2,
\vec \xi_3,\vec \xi_4)\, .
\label{annihirelr}
\end{eqnarray}
Here note has been taken of the fact that only the
component of the spatial part of the $\Delta(1232)$
wave function with the mixed symmetry 
$[31]_X$ that corresponds to the Young tableau:
\begin{equation}
{\young(123,4)}\, ,
\label{y1}
\end{equation}
contributes.
The coordinate vector $\vec \xi_3$ realizes this symmetry. 
With the explicit harmonic oscillator wave functions
(\ref{harmosc}) for the radial wave functions with the
arguments $\vec\xi_1\, ,..\,\vec\xi_4$, the matrix element takes the
form:
\begin{equation}
\langle T \rangle = 4 A_{p3}A_{\Delta 5}^{(J)}
C_C C_{FS} ({\omega_3\omega_5\over \omega^2})^3\, 
\sqrt{2}\omega_5 ({2\omega_5\over \sqrt{5\pi}})^3 
 \int d^3 \xi_3 {i\vec \xi_3\over \sqrt{3}}
e^{-2\sqrt{3} i \vec q_\pi \cdot \vec\xi_3/5}
e^{-4\xi_3^2\omega_5^2/5}\langle T_{45}\rangle\, .
\label{expl}
\end{equation}
In this expression oscillator wave functions of the form
(\ref{harmosc}), but with $\omega_3$ in place of
$\omega_5$, have been employed of the $qqq$ component of
the proton. The product of the factor 
$(\omega_3\omega_5/\omega^2)^3$, where
$\omega=\sqrt{(\omega_3^2+\omega_5^2)/2}$,
 and the coefficient $C_C$ is the overlap between the
antisymmetric color state $[111]_C$ of the proton in the
$qqq$ configuration and the first three components of the
mixed symmetry color state $[211]_C$ of the $qqqq\bar q$
component of the $\Delta(1232)$ in the color
configuration that is conjugate to (\ref{y1}). The
coefficient $C_{FS}$ is the corresponding overlap between
the mixed symmetry $[21]_{FS}$ flavor-spin state
of the proton in the $qqq$ configuration and the
corresponding $[4]_{FS}[22]_F[22]_S$ flavor-spin state
of the $qqqq\bar q$ component of the $\Delta(1232)$. 
These coefficients take the values:
\begin{equation} 
C_C= 1\, , \quad C_{FS} ={1\over \sqrt{6}}(\delta_{J1}
+{3\over \sqrt{5}}\delta_{J2})\, ,
\end{equation}
when the overall normalization factors of the color, space
and flavor-spin wave functions are included. 

The radial integral in (\ref{expl}) may be approximately 
evaluated with a power series expansion in $\vec q$ with
the result:
\begin{equation}
\langle T \rangle \simeq 4i A_{p3} A_{\Delta 5}^{(J)}
C_C C_{FS} ({\omega_3\omega_5\over \omega^2})^3\,
{q_+ \over \omega_5}{\sqrt{2}\over 4}
(1-{3\over 20}{q^2\over \omega_5^2})\,.
\label{ann2}
\end{equation}

Consider now the decay $\Delta^{++}(s_z=3/2)
\rightarrow p(s_z= 1/2)\,\pi^+$, where the $\Delta^{++}$
is in a $uuud\bar d$ and the proton in the $uud$
configuration.
In this case
one of the $u$ quarks annihilates
the $\bar d$ antiquark in the  $\Delta^{++}$ to
form the $\pi^+$. The complete 
amplitude for this annihilation process then 
becomes:
\begin{equation}
T= i{\sqrt{6}\over 3}
A_{p3} (A_{\Delta 5}^{(1)} + {3\over \sqrt{5}}
A_{\Delta 5}^{(2)})
({m g_A\over f_\pi})\,({\omega_3\omega_5\over \omega^2})^3\,
{q_+\over \omega_5}\,
(1-{3\over 20}{\vec q\,^2\over \omega_5^2})\,.
\label{diranni}
\end{equation}
This magnitude of this amplitude should then be
compared to that of the basic decay amplitude in
the pure 3-quark configuration (2), multiplied
by the factor $A_{p3}A_{\Delta 3}$. 
Note that in this case the phase of the $qqqq\bar q$
components (i.e. the signs of $A_{\Delta 5}^{(J)}$)
determines whether there will be constructive or
destructive interference with the decay
amplitudes for transitions between the $qqq$ or 
the $qqqq\bar q$ amplitudes without 
pair annihilation.

The annihilation mechanism will contribute to the
decay width of the $\Delta(1232)$ even in the
absence of $qqqq\bar q$ component in the proton.
Assume for the sake of an example 
that $A_{\Delta 5}^{(2)} = 0.32$, but
now that
$A_{p3}=1$ and $A_{\Delta 3}=0.95$. 
This implies a pure $qqq$ proton and a $\Delta^{++}$
with a 10\% probablility for the $qqqq\bar q$ 
configuration.
With
$m=340 $ MeV and $\omega_5 = 245$ MeV (\ref{omega}), 
it then follows
from these expressions that the direct
$qqqq\bar q \rightarrow qqq + \pi^+$  annihilation
mechanism, combined with the amplitude from the 
transition between $qqqq\bar q$ components, 
increases the calculated decay width
by $\sim$ 69\%.

The estimate above was based on the assumption that the
mean square radii of the $qqqq\bar q$ and $qqq$
components in the proton are equal. If the radius
of the $qqqq\bar q$ component is increased by
22 \% so that $\omega_5$ is reduced to 200 MeV, the
contribution from the direct annihilation
process (\ref{diranni}) is increased to an enhancement
of 81\%. 

\subsection{Confinement triggered annihilation}

\begin{figure}[t]
\vspace{20pt} 
\begin{center}
\mbox{\epsfig{file=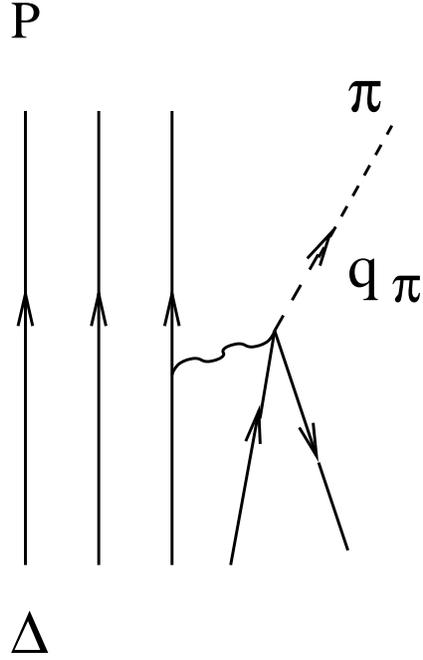, width=55mm}} 
%\mbox{\epsfig{file=f1,width=55mm}} 
\caption{Confinement induced
 $qqqq\bar q\rightarrow qqq\pi$ annihilation
diagram}
\label{fig2}
\end{center}
\vspace{10pt} 
\end{figure}

In addition to the direct annihilation mechanism that
is illustrated in Fig.\ref{fig1} the annihilation process
may also be triggered by the interaction between quarks.
The most obvious such triggered annihilation process
is that, which is caused by the confining interaction,
and which is illustrated in Fig.\ref{fig2}. 

The corresponding annihilation amplitude may be derived in 
the same way as the amplitude for the
direct annihilation process by inserting a quark propagator
multiplied by the confining interaction before and after the
pseudovector
pion-quark vertex. If only the point coupling and
pair terms are
retained after application of the Dirac equation, the
confinement triggered annihilation amplitude may, in the
case of a simple linear scalar confining interaction, be derived
from the direct annihilation amplitude by making the
substitution: 
\begin{equation}
m\rightarrow m+{1\over 2}(c r_{ij}-b) \,,
\label{shift}
\end{equation}  
where c is the string tension, $r_{ij}$ is the
distance between the two quarks that interact by the
confining interaction and b is a positive constant, which
makes the effective linear confining interaction potential
$cr-b$ negative at short distances. The presence of the $b$ term
is suggested by the phenomenology of the charm 
meson spectra \cite{timoH}.  
If the confining interaction
has the color coupling $\vec\lambda_i^C\cdot
\vec\lambda_j^C$, the string tension for $qq$ and
$q\bar q$ pairs in $qqqq\bar q$ systems
system is the same and equals half of the value
of
that for $qq$ pairs in three-quark systems \cite{genovese}. 

The substitution (\ref{shift}) may be viewed as a
a mass correction, which is natural in the case
of a scalar coupled confining interaction. A related mass
shift does in the case of charmonium serve to bring the
calculated M1 transition rates into agreement with the
empirical values \cite{nyfalt}.

The amplitude for confinement triggered annihilation
involves integration over two of the relative
coordinates $\vec\xi_i$ (\ref{relative}). By the overall
antisymmetry it is sufficient to consider the 
annihilation amplitude in Fig.\ref{fig2}, in which
the confining interaction between the 3rd and 4th quarks
triggers the annihilation
of the 4th quark against the antiquark. In this case the
confining interaction $(c/2\vert \vec r_3 - \vec r_4\vert-b/2)$
enters the integrand in the matrix element. As
$\vert \vec r_3 - \vec r_4\vert$ is proportional
to $\vert\vec\xi_2 -\sqrt{2}\vec\xi_3\vert$ the
integrals over the relative coordinates $\vec\xi_2$ and
$\vec\xi_3$ have to be done numerically. The relative
coordinates $\vec\xi_2$ and $\vec\xi_3$ realize the
mixed symmetry configurations
\begin{equation}
{\young(124,3)}\,, \qquad {\young(123,4)}\, ,
\label{y2}
\end{equation}
respectively. These combine with the corresponding
mixed color symmetry configurations
\begin{equation}
{\young(13,2,4)}\, ,\qquad {\young(14,2,3)}\, ,
\label{y3}
\end{equation}
in the 5-quark component of the
wave function of the $\Delta(1232)$. 

The matrix element of the confinement contribution may
be expressed as (cf.(\ref{annihi} ))
\begin{eqnarray} 
&&\langle T_{conf} 
\rangle = 12 A_{p3}A_{\Delta 5}^{(J)}\,\int \Pi_{i=1}^5 d^3 r_i\,
\psi_{p3}(\vec r_1,\vec r_2,\vec r_3)\,e^{i\vec q_\pi\cdot
(\vec r_4+\vec r_5)/2}\,\nonumber\\
&&\{{c\over 2}\vert\vec r_3 -\vec r_4\vert-{b\over 2}\}
\delta(\vec r_4-\vec r_5)\,
\langle T_{3,45}\rangle \,
\psi_{\Delta 5}(\vec r_1,\vec r_2,
\vec r_3,\vec r_4,\vec r_5) \delta(\vec R)\, .
\label{annihic}
\end{eqnarray}
Here the operator $T_{3,45}$ describes the annihilation 
process in Fig.\ref{fig2}. The factor 12 on the rhs is 
the number of contributing similar processes.

The matrix element (\ref{annihic}) may be rewritten in more
explicit form as:
\begin{eqnarray}
&&\langle T_{conf} \rangle = {6} A_{p3}A_{\Delta 5}^{(J)} C_C
C_{FS}\,({\omega_3\omega_5\over \omega^2})^{3/2}\,
{\sqrt{6}\over 3} \sqrt{2}\omega_5
({2\over \sqrt{5}})^3
({\omega_5\over \sqrt{\pi}})^6
\int d^3\xi_2 d^3\xi_3 \nonumber\\
&&{i\vec\xi_3\over \sqrt{3}}\,
c\,\{\vert\vec\xi_2-\sqrt{2}\vec\xi_3\vert -{\sqrt{6}b\over 2 c}\}
\,\langle T_{3,45}\rangle 
\,e^{-\omega^2\xi_2^2}\,
e^{-\alpha^2\xi_3^2}\,e^{-i\beta\vec q\cdot\xi_3}\,.
\label{confexpl}
\end{eqnarray}
The coefficients $\alpha$ and $\beta$ are defined
as 
\begin{equation}
\alpha = {2\over \sqrt{5}}\omega_5\, ,\quad
\beta={2\sqrt{3}\over 5}\,.
\label{coeffab}
\end{equation}

\begin{figure}[t]
\vspace{20pt} 
\begin{center}
\mbox{\epsfig{file=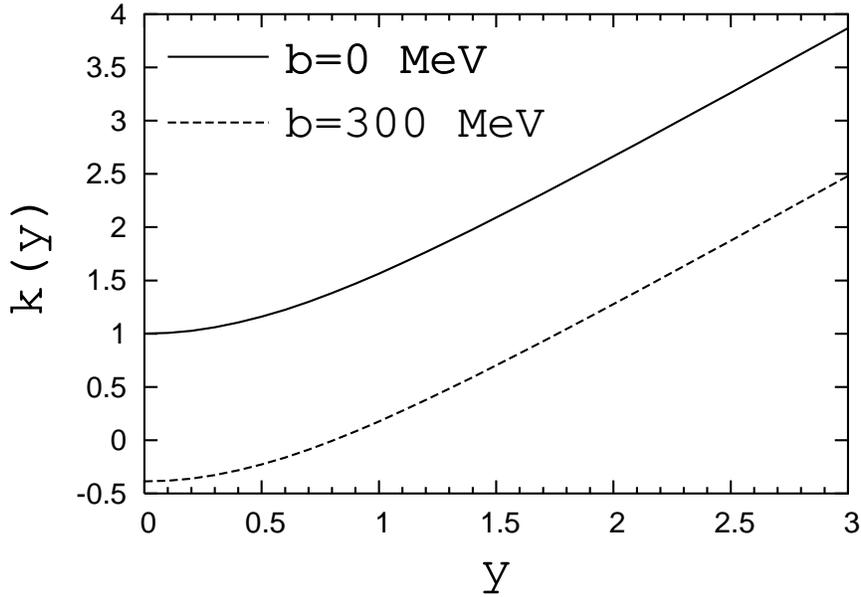, width=120mm}} 
\caption{The function k(y) with b = 0 and b = 300 MeV, $\omega_5$ = 245 MeV.}
\label{fig3}
\end{center}
\vspace{10pt} 
\end{figure}

The complete matrix element finally takes the form
\begin{equation}
T_{conf}=i({g_A\,c\over f_{\pi} \omega})\,({q_+\over\omega_5})\,
({\omega_3\omega_5\over \omega^2})^3\,
{512\sqrt{5}\over 125\pi}\,A_{p3}\,
(A_{\Delta 5}^{(1)}+{3\over\sqrt{5}}A_{\Delta 5}^{(2)})\,
K(q_\pi)\, .
\label{confexp2}
\end{equation}
Here the function $K(q)$ is defined as
\begin{equation}
K(q)=\omega_5^5\int_0^\infty d\xi_3\,\xi_3^4\,{j_1(\beta q\xi_3)
\over \beta q\xi_3}\,e^{-\alpha^2\xi_3^2}\, k(\omega\xi_3)\,,
\label{K(xi)}
\end{equation}
where $j_1$ is the spherical Bessel function of order 1
and the function $k(y)$ is defined as:
\begin{equation}
k(y)=\int_0^\infty dx x^2 e^{-x^2}\int_{-1}^1 dz\{\sqrt{x^2
-2\sqrt{2} xzy+2y^2}-{\sqrt{6}\over 2}{b\omega \over  c}\}\, .
\label{k(y)}
\end{equation}
With $b=0$ MeV, this function takes the value 1 at $y=0$ and approaches
the straight line $\sqrt{2\pi}y/2$ when $y>1$ as shown
in Fig.\ref{fig3}. For other values of the parameter $b$ the
function is shifted by a constant as also shown in the figure. 

\begin{figure}[t]
\vspace{20pt} 
\begin{center}
\mbox{\epsfig{file=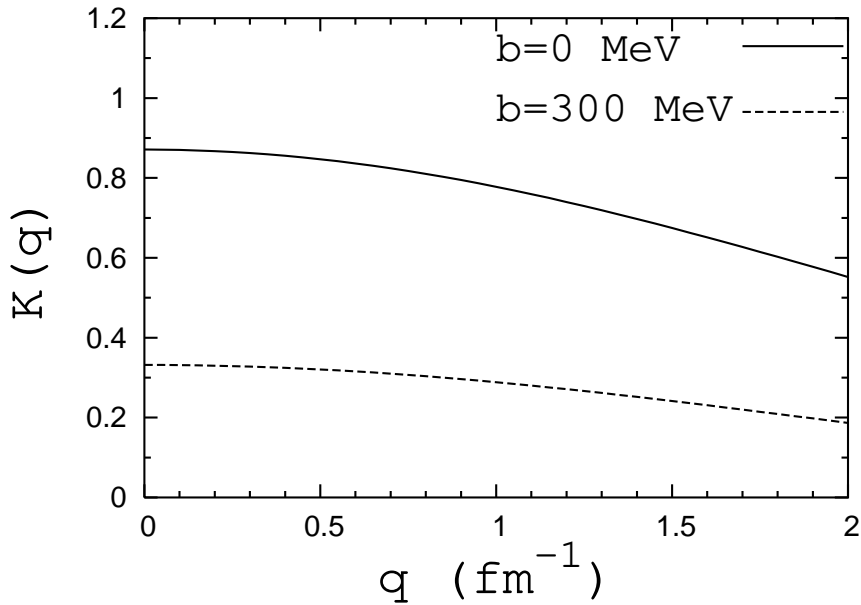, width=120mm}} 
\caption{The function K(q) with b = 0 and b = 300 MeV, $\omega_5$ = 245 MeV.}
\label{fig4}
\end{center}
\vspace{10pt} 
\end{figure}

The function $K(q)$ with $b=0$ and $300$ MeV is shown in Fig.\ref{fig4}. 
For $\Delta(1232)
\rightarrow N\pi$ $q_\pi= 227$ MeV and $K(q_\pi)=0.74$, when $b = 0$.

\begin{figure}[t]
\vspace{20pt} 
\begin{center}
\mbox{\epsfig{file=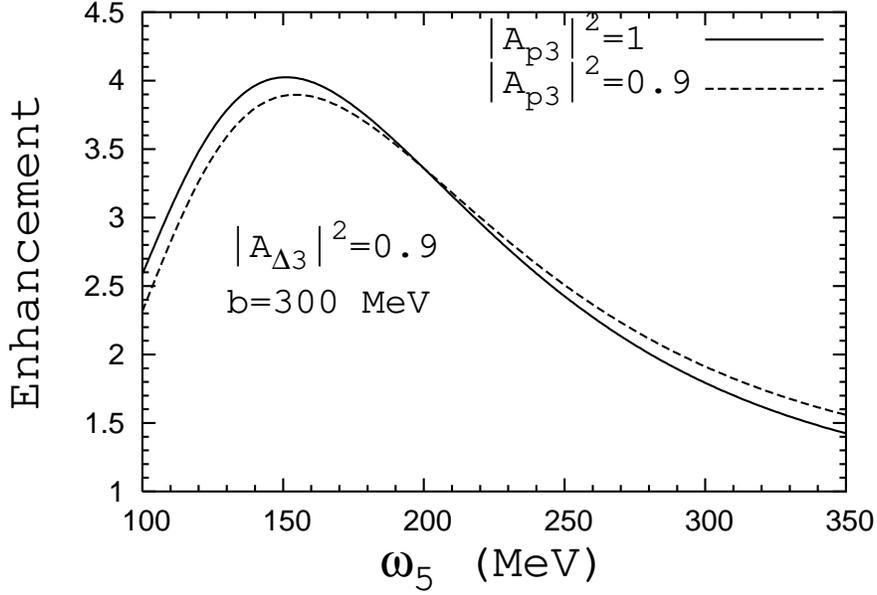, width=120mm}} 

\caption{The enhancement of the
calculated decay width as a function of the
oscillator parameter $\omega_5$ for the
$qqqq\bar q$ component of the $\Delta(1232)$ wave function
with the shift of the linear confining potential $b$ = 300 MeV. The
amplitudes for this component of the proton and
the $\Delta^{++}$ wave function are denoted
$A_{p3}$ and $A_{\Delta3}$ respectively. }
\label{fig5}
\end{center}
\vspace{10pt} 
\end{figure}

\begin{figure}[t]
\vspace{20pt} 
\begin{center}
\mbox{\epsfig{file=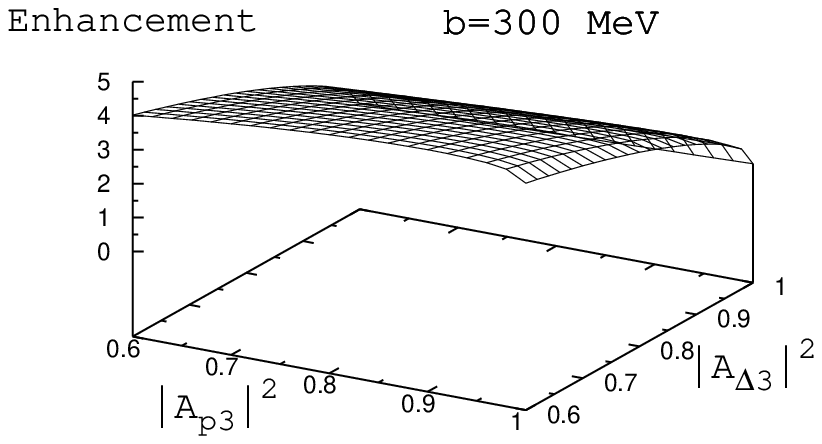, width=120mm}} 

\caption{The enhancement of the
calculated decay width as a function of the
amplitudes of the
$qqq$ components of the proton and the
the $\Delta^{++}$ wave functions
$A_{p3}$ and $A_{\Delta3}$. Here  $b$ = 300 MeV and the 
two oscillator parameters are
$\omega_3$ = 225 MeV and $\omega_5$= 245 MeV. }
\label{fig6}
\end{center}
\vspace{10pt} 
\end{figure}

For a numerical estimate of the significance of
confinement triggered annihilation
the string tension $c$ may be taken to one quarter of that
in $q\bar q$ systems \cite{genovese}. With a typical value
for that as $c_{q\bar q} = 1.12$ GeV/fm
\cite{timo2}, the value for the string tension in the
$qqqq\bar q$ system would be $c= 280$ MeV.
Consider again the previous example,
in which $A_{p3}=1$ and $A_{\Delta 5}^{(2)}=0.32$, but
now with the value 300 MeV for the shift parameter
$b$ in the confining potential. This value is chosen so as
to cover the range of values that have been used in
charm meson spectroscopy \cite{timoH}. The range
of values for $b$ between 0 and 300 MeV 
When the confinement triggered amplitude is
added to the amplitude for direct annihilation and
the amplitude for $\Delta^{++}\rightarrow p\pi^+$
decay in the appropriately normalized $qqq$ configuration 
it is found that the net effect is an increase by
a factor 2.5 of the decay width that is obtained in the $qqq$
configuration.
This estimate is based on the assumption that
the probability of the $qqqq\bar q$ component in the
$\Delta^{++}$ is 10\% and that the proton is a pure
$qqq$ state. The dependence of the
calculated enhancement 
on the oscillator parameter of the
$qqqq\bar q$
component of the $\Delta^{++}$ with $b$ = 300 MeV 
is shown in Fig.\ref{fig5}.
The enhancement as a function
of the amplitude of the $qqq$ components in the
proton and the $\Delta^{++}$ wave functions
is shown in Fig.\ref{fig6}. The dependence of the enhancement
of the decay width on the values of the shift parameter
$b$ in the
linear confining potential is given in Table \ref{tab:enhance}, 
from where it can be seen that, within a realistic range of the 
values of $b$ and with a 10\% $qqqq\bar q$ component in $\Delta(1232)$,
the final enhancement falls within the
range $2\sim 3$, which is substantial enough 
to compensate the underpredicted decay width of the pion 
decay of $\Delta(1232)$ in the $qqq$ quark model.

To have an estimate of the theoretical
uncertainty of the magnitude of the
contribution of the confinement triggered
annihilation process this estimate may be compared
to that, which is obtained if the linear confining interaction
between the quarks in the $qqqq\bar q$ system is 
replaced by the harmonic oscillator potential, which
corresponds to the wave function model employed above.
This is obtained by the substitution:
\begin{equation}
cr-b\rightarrow {1\over 2}C r^2-B\, ,
\label{osc}
\end{equation}
Here $B$ is a constant that shifts the interaction potential 
to negative values at short range. The oscillator constant 
$C$ is given as \cite{helminen}:
\begin{equation}
C = {m\omega_5^2\over 5}\,.
\label{bigC}
\end{equation}
With $m=340$ MeV and $\omega_5=245$ MeV this gives for
$C$ the value 105 MeV/fm$^2$.

\begin{figure}[t]
\vspace{20pt} 
\begin{center}
\mbox{\epsfig{file=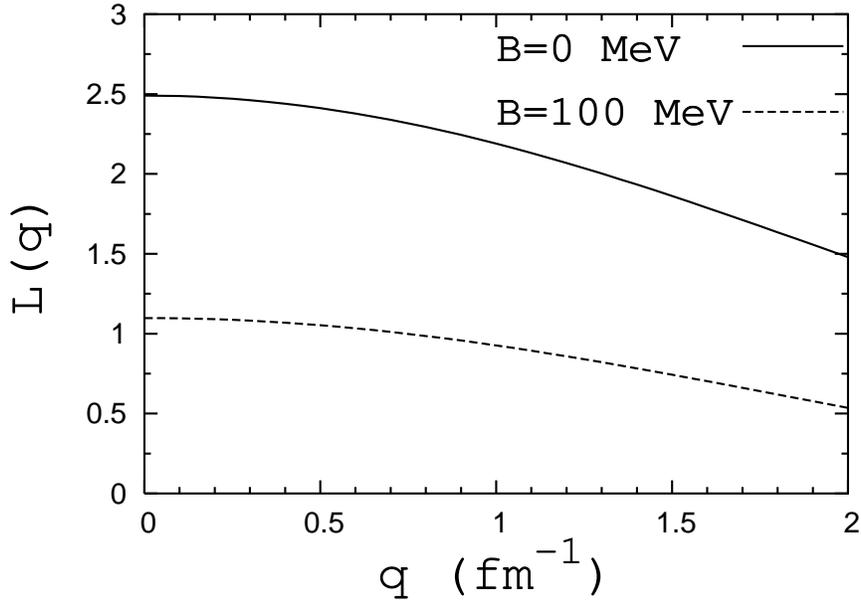, width=120mm}} 
\caption{The function L(q) with B = 0 and B = 100 MeV, $\omega_5$ = 245 MeV.}
\label{fig7}
\end{center}
\vspace{10pt} 
\end{figure}

The amplitude for confinement triggered annihilation
$\Delta^{++}\rightarrow p\pi^+$ in this oscillator
confinement model may be obtained directly from
the expression for linear confinement above
(\ref{confexp2}) by the substitution
\begin{equation}
c K(q_\pi)\,\rightarrow \, {\sqrt{6}\over 6}{C\over \omega}
L(q_\pi)\,.
\end{equation}
The function $L(q)$ is defined as the integral
\begin{equation}
L(q) = \sqrt{\pi}\omega_5^5\, \int_0^\infty d\xi_3\,
\xi_3^4\,{j_1(\beta  q \xi_3)\over \beta q\xi_3}\,
e^{-\alpha^2\xi_3^2}\,\{{3\over 4}+\omega^2(\xi_3^2-{3B\over 2C})\}\,.
\label{L(xi)}
\end{equation}
This function is plotted in Fig.\ref{fig7} for $B$ = 0 and 100 MeV. 
For $\Delta(1232)
\rightarrow N\pi$ decay $q_\pi= 227$ MeV and $L(q_\pi)=2.1$, when
B = 0.

With the numerical parameter values given above, the
magnitude of the confinement triggered annihilation
amplitude is smaller by a factor $\sim 2.7$ in the
model with oscillator confinement with $B$ = 0 than in
the case of the 
linear confining interaction with $b$ = 0 MeV. With the 
oscillator model with $B$ = 100 MeV,
the confinement triggered annihilation, when combined
with the amplitude for direct annihilation would lead
to an enhancement of the total calculated pion decay
width by a factor $\sim 2$. In Table \ref{tab:enhance} 
we list the calculated enhancement from the $qqq$ quark model 
value for different values of $B$ in the harmonic 
confining potential.   

\begin{table}
\caption{Calculated enhancement of the decay width value
in the the $qqq$ quark  model ($\delta$) for different values 
of the parameter $b$ in 
the linear confinement and $B$ in the harmonic confining potential.
Here the probability of the $qqqq\bar q$ component in the nucleon 
is taken to be zero and in the $\Delta(1232)$ 10\% and the oscillator
parameters $\omega_3$ = 225 MeV and $\omega_5$ = 245 MeV.}
\label{tab:enhance}
\begin{tabular}{c|ccccccc}
\hline 
   $b~~(MeV)$  & 150 & 200 & 250 & 300 & 350 & 400 &450\\
\hline
  $\delta$  & 3.30 & 3.03 & 2.76 & 2.51 & 2.27 &2.24&1.82\\ \hline
  \hline 
   $B~~(MeV)$  & 50 & 75 & 100 & 125 & 150 & 175&200\\
\hline
  $\delta$  & 2.23 & 2.12 & 2.00 & 1.90 & 1.79 &1.69&1.59\\ \hline  
\end{tabular}
\end{table}

These results show that there is a considerable model dependence
in the calculated enhancement of the decay with that arises
from annihilation transitions that are triggered by the
confining interaction. The results thus have to be viewed
as qualitative and of a very exploratory nature.

\section{Discussion}
\label{disc}

Above it was shown that quark-antiquark annihilation
may contribute significantly to the decay
width of the $\Delta(1232)$ resonance calculated
in the non-relativistic constituent quark model. This contribution
depends on the amplitude of the $qqqq\bar q$ admixture of
the $\Delta(1232)$ and on the nature of the interaction
between the quarks and the antiquark. 
The enhancement of the decay width calculated in the
valence quark model was found to be as large 
as factors 2 -- 3 if the $\Delta(1232)$ contains a $qqqq\bar q$
component with 10\% probability. 
This large contribution was obtained
with the assumption that the Lorentz nature of the
confining interaction is a scalar interaction. This is sufficient
to compensate for the underestimate of the decay
width of the $\Delta(1232)$ in the $qqq$ valence
quark model.

The present estimates relied on a non-relativistic
harmonic oscillator
model for the quark wave functions, which has previously
been shown to provide useful, if qualitative, information
on baryon phenomenology. To go beyond this model
requires a detailed model for the interaction between
the quarks, which should be constrained both by the
empirical splitting between the $\Delta(1232)$
and the nucleon as well as by the electromagnetic form
factors of the nucleon. Only with such a Hamiltonian model
is it possible to obtain quantitative estimates for the
amplitude of the $qqqq\bar q$ components in the baryons.
A more quantitative calculation should also require
covariant framework. The fact that the covariant
quark models with instant form kinematics lead to rather 
similar results as the non-relativistic quark model
\cite{coester}, 
suggests that the qualitative features of the present
results will carry over to covariant descriptions
based on instant form kinematics.

It should be instructive to extend this phenomenological
analysis to the case of the low lying positive parity
resonances $N(1440)$ and the $\Delta(1600)$, which
are very likely to have substantial sea-quark components.
The widths of these are typically underestimated by
large factors in calculations based on the 3-valence
quark model \cite{bruno, melde}.

\vspace{0.2cm}
\begin{acknowledgments}

Research supported in part by the Academy of Finland grant 
number 54038 

\end{acknowledgments}


\begin{thebibliography}{99}

\bibitem{bruno} B. Juli\'a-D{\'i}az, D. O. Riska and
F. Coester, Phys. Rev. {\bf C70}, 045205 (2004)

\bibitem{melde} T. Melde et al., Phys. Rev. {\bf C72}, 015207 (2005)      

\bibitem{sato} T. Sato and T.-S. H. Lee, Phys. Rev.
{\bf C63}, 055201 (2001)

\bibitem{sato2} T. Sato, D. Uno and T.-S. H. Lee,
Phys. Rev. {\bf C67}, 065201 (2003)

\bibitem{towell} R. S. Towell et al., Phys. Rev. {\bf D64}
052002 (2002)

\bibitem{NA51} A. Baldit et al., Phys. Lett. {\bf B332},
244 (1994)

\bibitem{NMC} P. Amaudruz et al., Phys. Rev. Lett.
{\bf 66}, 2712 (1994) 

\bibitem{HERMES} K. Ackerstaff et al., Phys. Rev. Lett.
{\bf 81}, 5519 (1998)

\bibitem{helminen} C. Helminen and D. O. Riska, 
Nucl. Phys. {\bf A699}, 624 (2002) 

\bibitem{glozman} L. Ya. Glozman and D. O. Riska,
Phys. Rept. {\bf 268}, 263 (1996)

\bibitem{close} F. Close, An Introduction to Quarks and
Partons, Academic Press, London (1979)

\bibitem{chen} J.-Q. Chen, Group Representation Theory for
Physicists, World Scientific, Singapore (1989)

\bibitem{brunomag} B. Juli\'a-D{\'i}az and D. O. Riska,
Nucl. Phys. {\bf A739}, 69 (2004)

\bibitem{garvey} G.T. Garvey and J.C. Peng, Prog. Part. Nucl.
Phys. {\bf 47}, 203 (2001)

\bibitem{timoH} K. O. Henriksson et al., Nucl. Phys. {\bf A686},
355 (2001)

\bibitem{genovese} M. Genovese, J. -M. Richard, Fl. Stancu
and S. Pepin, Phys. Lett. {\bf B425}, 171 (1998)

\bibitem{nyfalt} T. A. L\"{a}hde, C. Nyf\"{a}lt and D. O. Riska,
Nucl. Phys. {\bf A645} (1999) 587

\bibitem{timo2} T. A. L\"{a}hde and D. O. Riska,
Nucl. Phys. {\bf A693} (2001) 755

\bibitem{coester} F. Coester and D. O. Riska,
Nucl. Phys. {\bf A728}, 439 (2003)

\end{thebibliography}
\end{document}